\newcommand{\gevm}{~GeV/$c^2$}
\newcommand{\phione}{$\phi^{}_1$}
\newcommand{\phitwo}{$\phi^{}_2$}
\newcommand{\belle}{Belle}

\newcommand{\cp}{$CP$}
\newcommand{\mbc}{$m^{}_{\rm bc}$}
\newcommand{\deltaE}{$\Delta E$}

\newcommand{\cpipi}{${\cal C}^{}_{\pi\pi}$}
\newcommand{\spipi}{${\cal S}^{}_{\pi\pi}$}
\newcommand{\crhopi}{${\cal C}^{}_{\rho\pi}$}
\newcommand{\srhopi}{${\cal S}^{}_{\rho\pi}$}

\newcommand{\bbar}{\overline{B}{}^{\,0}}
\newcommand{\bbbar}{$B^0$-$\bbar$}

\newcommand{\ra}{\!\rightarrow\!}

\newcommand{\bpipi}{$B^0\ra\pi^+\pi^-$}
\newcommand{\brhopi}{$B^0\ra\rho^\pm\pi^\mp$}

\def\simge{\mathrel{%
   \rlap{\raise 0.511ex \hbox{$>$}}{\lower 0.511ex \hbox{$\sim$}}}}

\def\plusominus{\mathrel{%
   \rlap{\raise 0.651ex \hbox{\footnotesize $+$}}
   {\lower 0.651ex \hbox{\footnotesize $-$}}}}
\def\minusoplus{\mathrel{%
   \rlap{\raise 0.651ex \hbox{\footnotesize $-$}}
   {\lower 0.651ex \hbox{\footnotesize $+$}}}}

\documentclass{preprint-ijmpa}

\begin{document}

\markboth{A.\,J.\,Schwartz}
{Constraints upon the CKM angle \phitwo}

\catchline{}{}{}{}{}
%
%%%%%%%%%%%%%%%%%%%%%%%%%%%%%%%%%%%%%%%%%%%%%%%%%%%%%%%%%%%%%%%%%%%%

\title{
\vskip-0.50in
\begin{flushright}
{\normalsize UCHEP-04-07}
\end{flushright}
\vskip0.40in
Constraints upon the CKM angle \phitwo\ ($\alpha$) from \belle }

\author{\footnotesize A.\ J.\ Schwartz}

\address{Physics Department, University of Cincinnati, 
P.O.\ Box 210011, Cincinnati, Ohio 45221 USA}

\author{(For the Belle Collaboration)}

\maketitle

%\pub{Received (1 November 2004)}
%{Revised (Day Month Year)}

\begin{abstract}
The \belle\ experiment has measured branching 
fractions and \cp\ asymmetries for the charmless decays 
\bpipi\ and $B^0\ra\rho^\pm\pi^\mp$. From these measurements, 
constraints upon the CKM angle \phitwo\ can be obtained. These 
constraints indicate that \phitwo\ is around~100$^\circ$.
%\keywords{mixing; doubly-Cabibbo-suppressed; charm decays.}
\end{abstract}

\section{Overview}

The Standard Model predicts \cp\ violation to occur in $B^0$
meson decays owing to a complex phase in the $3\times 3$
Cabibbo-Kobayashi-Maskawa (CKM) mixing matrix. This phase
is illustrated by plotting the unitarity condition
$V^*_{ub} V^{}_{ud} + V^*_{cb} V^{}_{cd} + V^*_{tb} V^{}_{td} =0$ 
as vectors in the complex plane: the phase results in 
a triangle of nonzero height.  One interior angle of the 
triangle, denoted \phione\ or $\beta$, is determined from 
$B^0\ra J/\psi\,K^0$ decays.\footnote{Charge-conjugate modes 
are included throughout this paper unless noted otherwise.}
Another interior angle, \phitwo\ or $\alpha$, is determined 
from charmless decays such as \bpipi\ and \brhopi. Here we 
present measurements of these charmless decays from 
\belle,\cite{belle} and the resulting constraints 
upon \phitwo. 

To select $B^0\ra\pi^+\pi^-/\rho^\pm\pi^\mp$ decays, 
we require two opposite-charge pion-candidate tracks 
originating from the interaction region. For \brhopi, 
one of these tracks is combined with a $\pi^0$ candidate. 
The charged pion identification criteria are based on 
information from time-of-flight counters, aerogel 
cherenkov counters, and $dE/dx$ information from 
the central tracker.\cite{bellePID} 
$B$ decays are identified via 
the ``beam-constrained'' mass
$m^{}_{\rm bc}\equiv\sqrt{E^2_b-p^2_B}$ and 
the energy difference $\Delta E\equiv E^{}_B-E^{}_b$,
where
$p^{}_B$ is the reconstructed $B$ momentum, 
$E^{}_B$ is the reconstructed $B$ energy,
and $E^{}_b$ is the beam energy, all evaluated in 
the $e^+e^-$ center-of-mass (CM) frame.
The \mbc\ and \deltaE\ distributions
are jointly fit for the signal event yields. 

A tagging algorithm is used to identify the flavor of the 
$B$ decay, i.e., $B^0$ or~$\bbar$.
This algorithm examines tracks not associated with the 
signal decay to identify the flavor of the non-signal~$B$. 
The signal-side tracks are fit for a decay vertex, and the 
tag-side tracks are fit for a separate decay vertex; the 
distance $\Delta z$ between vertices is (to good approximation) 
proportional to the time difference between the $B$ 
decays: $\Delta z\approx (\beta\gamma c)\Delta t$, where
$\beta\gamma$ is the Lorentz boost of the $e^+e^-$ system.
%and equals 0.43.

The dominant background for both modes is
$e^+e^-\ra q\bar{q}$ continuum events, where
$q=u,d,s,c$. In the CM frame such events 
tend to be collimated along the beam 
directions, whereas $B\overline{B}$ events 
tend to be spherical. The ``shape'' of 
an event is quantified via Fox-Wolfram 
moments\cite{FoxWolfram} of the form
$h^{}_\ell = \sum_{i,j} p^{}_i\,p^{}_j\,P^{}_\ell(\cos\theta^{}_{ij})$,
where $i$ runs over all tracks on the tagging side and
$j$ runs over all tracks on either the tagging side or
the signal side. The function $P^{}_\ell$ is the $\ell$th
Legendre polynomial and $\theta^{}_{ij}$ is the angle 
between momenta $\vec{p^{}_i}$ and $\vec{p^{}_j}$ in
the CM frame. These moments are combined into a Fisher 
discriminant, and this is combined with the probability
density function for the cosine of the angle
between the $B$ direction and the electron beam
direction. This yields an overall likelihood~${\cal L}$. 
Continuum events are rejected by requiring that the ratio
${\cal L}_{B\overline{B}}/
({\cal L}_{B\overline{B}}+ {\cal L}_{q\bar{q}})$
be greater than a minimum value.

\section{{\boldmath \bpipi}}

The decay time dependence of $B^0/\bbar\ra\pi^+\pi^-$ 
decays is given by\cite{gronau}
\begin{eqnarray}
\frac{dN}{d\Delta t} & \ \propto\  & 
e^{-\Delta t/\tau}\biggl[1 - q\,{\cal C}^{}_{\pi\pi}\cos(\Delta m\Delta t)
+ q\,{\cal S}^{}_{\pi\pi}\sin(\Delta m \Delta t)\,\biggr], 
\label{eqn:pipi}
\end{eqnarray}
where $q\!=\!+1$ ($-1$) corresponds to $B^0$ ($\bbar$) 
tags, and $\Delta m$ is the \bbbar\ mass difference. The 
parameters \cpipi\ and \spipi\ are \cp-violating and 
related to \phitwo\ via\cite{gronaurosner}
\begin{eqnarray}
{\cal C}^{}_{\pi\pi} & \ =\ & \frac{1}{R}\cdot
\left( 2\left|\frac{P}{T}\right|\sin(\phi^{}_1 - \phi^{}_2)\sin\delta\right)
\label{eqn:cpipi} \\
 & & \nonumber \\
{\cal S}^{}_{\pi\pi} & \ =\  & 
\frac{1}{R}\cdot
\biggl( 2\left|\frac{P}{T}\right|\sin(\phi^{}_1-\phi^{}_2)\cos\delta +
 \sin\,2\phi^{}_2 - \left|\frac{P}{T}\right|^2\sin\,2\phi^{}_1 \biggr)
\label{eqn:spipi} \\
 & & \nonumber \\
R & \ =\  & 
1-2\left|\frac{P}{T}\right|\cos(\phi^{}_1+\phi^{}_2)\cos\delta + 
\left|\frac{P}{T}\right|^2\,,
\end{eqnarray}
where $\phi^{}_1=(23.2^{+1.6}_{-1.5})^\circ$,\cite{hfag}
$|P/T|$ is the magnitude of a possible penguin amplitude 
relative to that of the tree-level amplitude, and $\delta$ is 
the strong phase difference between the two amplitudes. If there 
were no penguin contribution, ${\cal C}^{}_{\pi\pi}=0$ and 
${\cal S}^{}_{\pi\pi} = \sin 2\phi^{}_2$.
Since Eqs.~(\ref{eqn:cpipi}) and (\ref{eqn:spipi}) have three
unknown parameters, measuring \cpipi\ and \spipi\ determines 
a volume in $\phi^{}_2$ - $\delta$ - $|P/T|$ space.

\begin{figure}[t]
\centerline{\epsfxsize=6.25cm \epsfysize=3.0cm \epsfbox{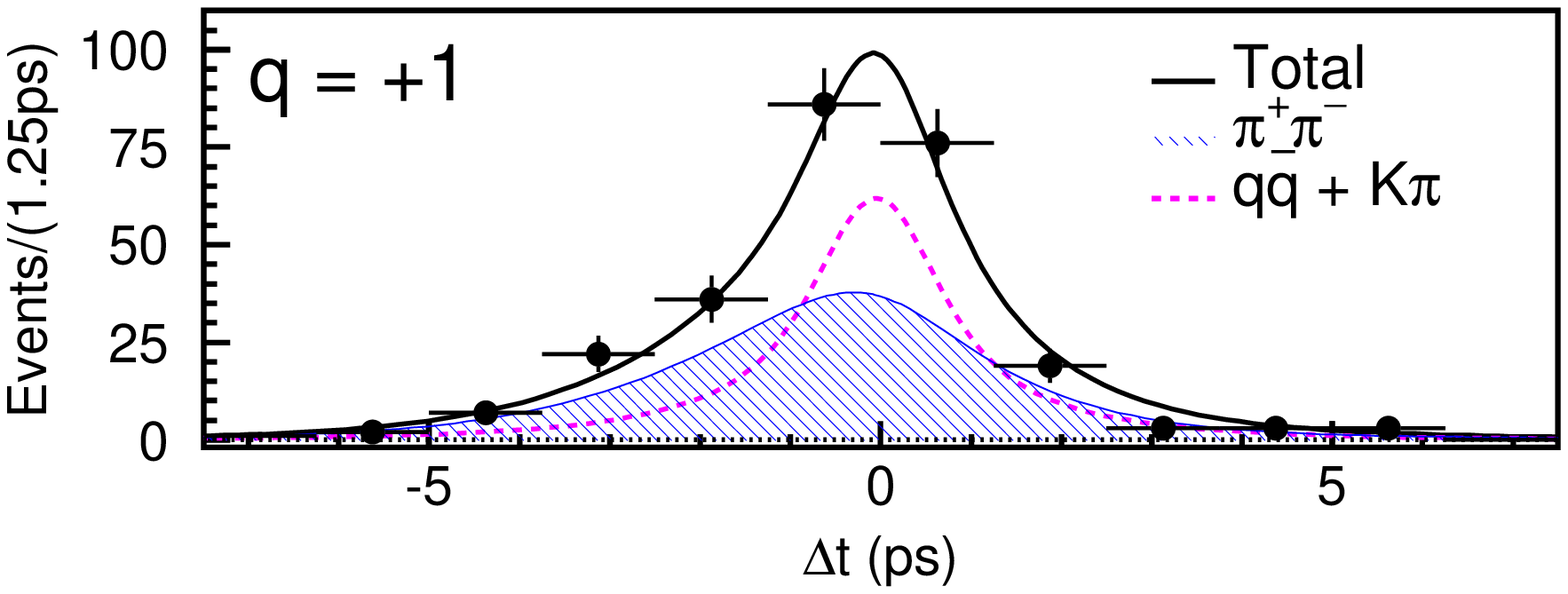}
\hskip0.05in
\epsfxsize=6.25cm \epsfysize=3.0cm \epsfbox{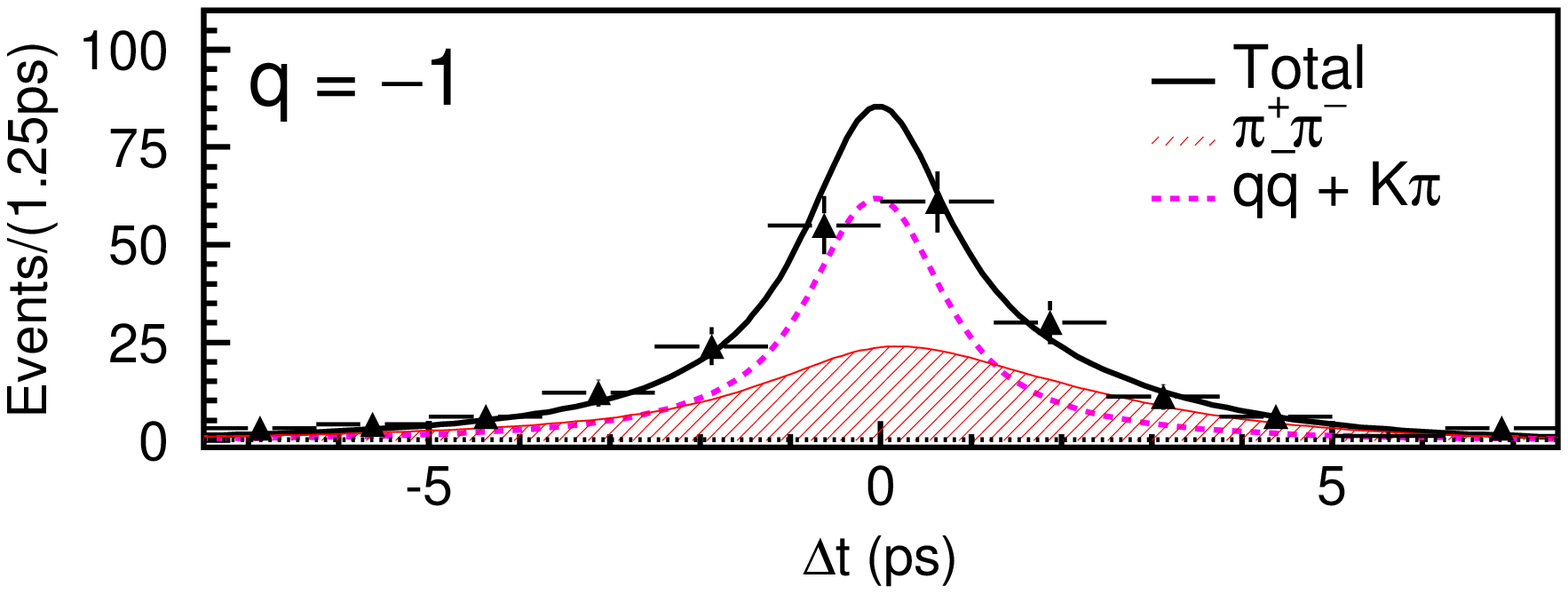}}
\caption{The \bpipi\ $\Delta t$ distribution for $q\!=\!+1$ 
tags (left) and $q\!=\!-1$ tags (right). The 
points show the data and the smooth curves 
are projections of the unbinned ML fit. }
\label{fig:pipi1}
\end{figure}

The most recent \belle\ measurement of \cpipi\ and \spipi\ 
is with 140~fb$^{-1}$ of data.\cite{belle_pipi} The event 
sample consists of 224 $\bbar\ra\pi^+\pi^-$ candidates and 
149 \bpipi\ candidates after background subtraction. 
These events are subjected to an unbinned 
maximum-likelihood (ML) fit in~$\Delta t$; the results are
${\cal C}^{}_{\pi\pi}\!=\!-0.58\,\pm\,0.15\,({\rm stat})\,\pm\,0.07\,({\rm syst})$ and
${\cal S}^{}_{\pi\pi}\!=\!-1.00\,\pm\,0.21\,({\rm stat})\,\pm\,0.07\,({\rm syst})$,
which indicate large \cp\ violation. The nonzero value for
${\cal C}^{}_{\pi\pi}$ indicates {\it direct\/} \cp\ violation.
Fig.~\ref{fig:pipi1} shows the $\Delta t$ distributions 
for $q=\pm 1$ tagged events; a clear difference is seen 
between the distributions. 

These values determine a 
95\% C.L. volume in $\phi^{}_2$ - $\delta$ - $|P/T|$ space.
Projecting this volume results in the constraints
$90^\circ < \phi^{}_2 < 146^\circ$ for $|P/T|<0.45$
(as predicted by QCD factorization\cite{qcd_fctrztn} 
and perturbative QCD\cite{pert_qcd}), and 
$|P/T|>0.17$ for any value of~$\delta$.
The dependence upon $|P/T|$ and $\delta$ can be 
removed by making additional theoretical assumptions. 
A model based on $SU(3)$ symmetry that uses the measured 
rates for $B^+\ra K^0\pi^+$ and $B^0\ra K^+\pi^-$ obtains 
$\phi^{}_2\!=\!(103\,\pm\,17)^\circ$.\cite{gronaurosner2}

\section{{\boldmath \brhopi}}

For \brhopi\ the final state is not a \cp\ eigenstate,
and there are four decays to consider: 
$B^0\ra\rho^\pm\pi^\mp$ and $\bbar\ra\rho^\pm\pi^\mp$. 
The rates can be parametrized as\cite{rhopi}
\begin{eqnarray}
\frac{dN(B\ra\rho^\pm\pi^\mp)}{d\Delta t} & \ \propto\  & 
(1\pm A^{\rho\pi}_{CP})\ \times  \nonumber \\
 & & \hskip-1.0in
e^{-\Delta t/\tau}\biggl[1 - q\,({\cal C}^{}_{\rho\pi}\pm 
\Delta {\cal C}^{}_{\rho\pi})\cos(\Delta m\Delta t)\ 
+\ q\,({\cal S}^{}_{\rho\pi}\pm \Delta {\cal S}^{}_{\rho\pi})
\sin(\Delta m \Delta t)\,\biggr], 
\label{eqn:rhopi}
\end{eqnarray}
where $q\!=\!+1$ ($-1$) corresponds to $B^0$ ($\bbar$) 
tags. The parameters \crhopi\ and \srhopi\ are \cp-violating, 
whereas $\Delta{\cal C}^{}_{\rho\pi}$ 
and $\Delta{\cal S}^{}_{\rho\pi}$ are \cp-conserving. 
$\Delta{\cal C}^{}_{\rho\pi}$ characterizes the 
difference in rates between the ``$W\ra\rho$'' process
$B^0\ra\rho^+\pi^-$ or $\bbar\ra\rho^-\pi^+$ and the
``${\rm spectator}\ra\rho$'' process
$B^0\ra\rho^-\pi^+$ or $\bbar\ra\rho^+\pi^-$.
$\Delta{\cal S}^{}_{\rho\pi}$ depends, in addition,
on differences in phases between the $W\ra\rho$ 
and ${\rm spectator}\ra\rho$ amplitudes.
The parameter $A^{\rho\pi}_{CP}$ is the time 
and flavor integrated asymmetry
$\Gamma(B^0\ra\rho^+\pi^-) + \Gamma(\bbar\ra\rho^+\pi^-)  
- \Gamma(\bbar\ra\rho^-\pi^+) - \Gamma(B^0\ra\rho^-\pi^+)$
divided by the sum of the four rates. We also define the
asymmetries
$A^{}_{\plusominus\minusoplus}\equiv
\left[N(\bbar\ra\rho^\mp\pi^\pm)-N(B^0\ra\rho^\pm\pi^\mp)\right]/
	\left[N(\bbar\ra\rho^\mp\pi^\pm)+N(B^0\ra\rho^\pm\pi^\mp)\right]$. 
$A^{}_{+-}$ depends only on $W\ra\rho$ processes and 
$A^{}_{-+}$ depends only on ${\rm spectator}\ra\rho$ processes.

\begin{figure}[t]
\centerline{\epsfxsize=9.8cm \epsfbox{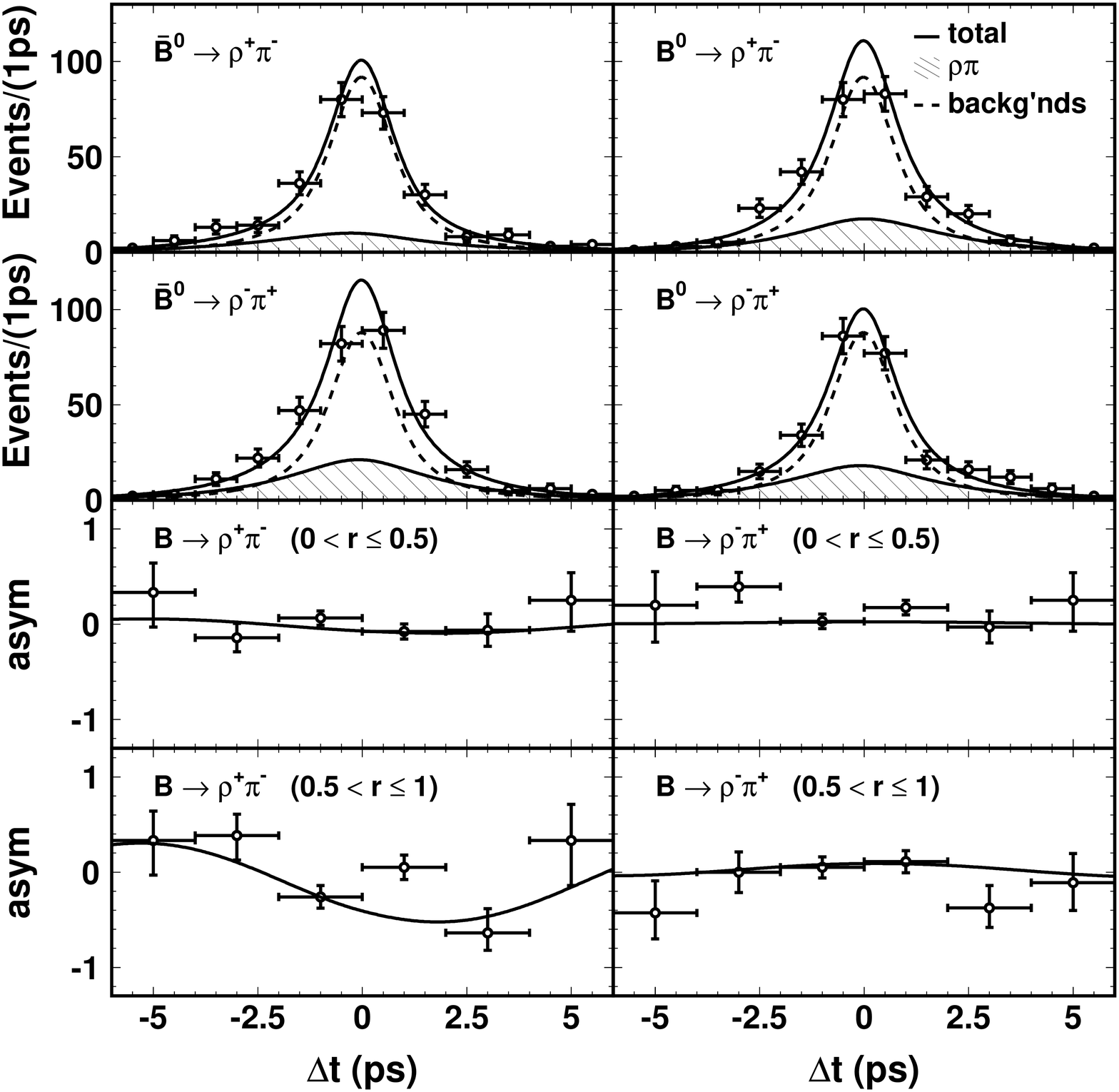}}
\caption{
The \brhopi\ $\Delta t$ distribution for $q\!=\!+1$ tags (left) 
and $q\!=\!-1$ tags (right), and the resulting \cp\ asymmetry 
(bottom). The asymmetry is shown separately for high-quality 
($r>0.5$) and low quality ($r\leq 0.5$) tags. The smooth 
curves are projections of the unbinned ML fit.}
\label{fig:rhopi2}
\end{figure}

The most recent \belle\ results are from 140~fb$^{-1}$ 
of data.\cite{belle_rhopi} To remove charge-ambiguous decays 
and possible interference between $B^0\ra\rho^+\pi^-$ and 
$B^0\ra\rho^-\pi^+$ amplitudes, we require both
$0.57\!<\!m^{}_{\pi^\pm\pi^0}\!<\!0.97$\gevm\ 
and $m^{}_{\pi^\mp\pi^0}\!>\!1.22$\gevm. 
We define a signal region 
$m^{}_{\rm bc}\!>\!5.27$\gevm\ and 
$-0.10\!<\!\Delta E\!<\!0.08~{\rm GeV}$;
there are 1215 events in this region and 329 \brhopi\ 
candidates. Fitting to $\Delta t$ yields
$A^{\rho\pi}_{CP}= -0.16\pm 0.10\pm 0.02$,
$C^{}_{\rho\pi} = 0.25\pm 0.17\,^{+0.02}_{-0.06}$,
$S^{}_{\rho\pi}= -0.28\pm 0.23\,^{+0.10}_{-0.08}$,
$\Delta C^{}_{\rho\pi}= 0.38\pm 0.18\,^{+0.02}_{-0.04}$,
and $\Delta S^{}_{\rho\pi} = -0.30\pm 0.24\pm 0.09$.
From these values we calculate
$A^{}_{+-} = -0.02\pm 0.16\,^{+0.05}_{-0.02}$ and
$A^{}_{-+} = -0.53\pm 0.29\,^{+0.09}_{-0.04}$.
The first errors listed are statistical and the
second systematic.
The $\Delta t$ distributions for $q=\pm 1$ tagged 
events and the resulting \cp\ asymmetry 
are shown in Fig.~\ref{fig:rhopi2}.

These values can be used to constrain \phitwo. 
However, since the penguin contribution is unknown, additional 
information is needed. An $SU(3)$-based model that uses the 
measured rates or limits for 
$B^0\ra K^{*\pm}\pi^\mp$, 
$B^0\ra\rho^\mp K^\pm$, 
$B^\pm\ra K^{*0}\pi^\pm$, and
$B^\pm\ra\rho^\pm K^0$ obtains
$\phi^{}_2=102\pm 19^\circ$.\cite{gronauzupan}
This value is surprisingly close to that 
resulting from the values of \cpipi\ and 
\spipi\ measured in \bpipi\ decays.

\end{document}